\documentclass[doublecol]{epl2} 
\title{Transparency at the interface between two isospectral crystals}
\shorttitle{Transparency at the interface ... } %Insert here a short version of the title if it exceeds 70 characters

\author{S. Longhi \inst{1,2} \and G. Della Valle \inst{1,2}}
\shortauthor{S. Longhi \etal}

\institute{                    
  \inst{1}  Dipartimento di Fisica, Politecnico di Milano, Piazza L. da Vinci 32, I-20133 Milano, Italy\\
  \inst{2}  Istituto di Fotonica e Nanotecnlogie del Consiglio Nazionale delle Ricerche, sezione di Milano, Piazza L. da Vinci 32, I-20133 Milano, Italy}

\abstract{Reflection at an interface separating two different media is a rather universal phenomenon which arises because of wave mismatching at the interface. By means of supersymmetric quantum mechanics methods, it is shown that a fully transparent interface can be realized, connecting two isospectral but different one-dimensional crystals. An example of reflectionless interface is presented for the sinusoidal (Mathieu) crystal connected to a non-sinusoidal potential by a transparent domain wall.}

\begin{document}

\maketitle

\section{Introduction}
Reflection, transmission and localization of matter or classical waves at the interface between two different media are fundamental phenomena encountered in different areas of physics. Interfaces are generally responsible for the appearance of surface waves that are localized at the boundary of the two media. For example, electronic surface states can be sustained at the interface between two different semiconductors in the gap between conduction and valence bands \cite{S,James}. Surface acoustic waves   can arise at the interface of two media owing to the coupling between longitudinal and transverse modes \cite{aco}.  Similarly, surface electromagnetic waves can be sustained at the interface between two different periodic optical crystals \cite{Yariv,cr1,cr2}, as well as at a metal-dielectric interface in the form of plasmon waves \cite{plasmon}. \par   
Other interesting properties of interfaces arise when considering reflection and transmission phenomena. Refection at an interface is a rather universal phenomenon which arises from the abrupt change of the medium properties. A well-known example is provided, for instance, by Fresnel reflection at the interface between two dielectric optical media. At an interface between a normal metal and a superconductor, Andreev reflection \cite{A1}, i.e. conversion of electron into hole excitations by the superconducting pair potential, takes place in the form of either retro-type or specular reflection \cite{A2}.  Andreev reflection and electron-electron retro-type reflection can be observed, for example, at the interface of graphene and superconductors \cite{A2,A3,A4}.   In some exceptional cases, an abrupt interface between two media can appear to be transparent (reflectionless interface). A remarkable example of transparency is provided by a potential step or barrier in graphene junctions \cite{A2,G1,G2,G3}, which turn out to be reflectionless at any energy of the incoming electron at normal incidence. In this case transparency is attributed to Klein tunneling of massless Dirac fermions in a gapless band system with linear dispersion relations of valence and conduction bands. 
Other kinds of transparency effects at interfaces have been predicted to occur in superconductor/ferromagnetic multilayers \cite{uff1}, 
ac-driven tight-binding lattices \cite{uff2}, transformation optical metamaterials \cite{uff3}, and topological insulators with induced
ferromagnetic domain walls \cite{uff4}. In the latter case transparency was related to a supersymmetry of the surface Dirac Hamiltonian
combined with the domain-wall profile.\par
 In this Letter we show rather generally that perfect transparency can be realized at an interface between two isospectral crystals. Isospectrality is a necessary, but not a sufficient condition, to realize a transparent interface {\it at any} energy \cite{note0}. The existence of crystals with distinct unit cells supporting the same band structure is possible owing to the non-uniqueness of the inverse  spectral problem for the Schr\"{o}dinger equation with a periodic potential. Isospectral crystals can be synthesized by supersymmetric (susy) quantum mechanics  and Darboux  transformation methods \cite{susy1,susy2,per}, which are powerful techniques in spectral problems. However, an interface between two isospectral crystals is generally not transparent, and the possibility to realize a reflectionless interface between two different isospectral crystals has not been investigated yet.  By a suitable extension of susy methods to a periodic lattice, we show that a fully transparent interface between two isospectral crystals, which is reflectionless at any energy, can be synthesized. An example of reflectionless interface is presented for a sinusoidal (Mathieu) periodic lattice, connected by a transparent domain wall to a non-sinusoidal lattice.

\section{Isospectral crystals} 
Let us consider two periodic potentials $V_1(x)$ and $V_2(x)$ describing two crystals with the same lattice period $a$ but with different unit cells, i.e. $V_{1,2}(x+a)=V_{1,2}(x)$ and $V_1(x) \neq V_2(x)$. The two crystals are said to be isospectral if they have the same band structure \cite{note}. Isospectral crystals can be synthesized, for instance, by application of the susy methods of quantum mechanics  \cite{susy1,susy2}. The standard procedure to realize isospectral crystals is discussed in several works (see, for instance, \cite{per}) and it is here briefly reviewed for the sake of clearness. \par Let us consider the Hamiltonian $\hat{H}_1=-d^2/dx^2+V_1(x)$ with a periodic potential $V_1(x)$ of period $a$, and let us assume without lack of generality that the energy $E=0$ falls in a gap of the spectrum of $\hat{H}_1$. Indicating by $\chi (x)$ an arbitrary solution to the equation $\hat{H}_1 \chi=0$, the following factorization holds
\begin{equation}
\hat{H}_1= \hat{A}^{\dag} \hat{A}
\end{equation}
where $\hat{A}=-d/dx+W(x)$ and $W(x)$ is the so-called superpotential, given by
\begin{equation}
W(x)=\frac{d}{dx} {\rm ln} \; [ \chi(x)]=\frac{1}{\chi(x)} \frac{d \chi}{dx}
\end{equation}
The potential $V_1(x)$ is related to the superpotential $W(x)$ by the simple relation
\begin{equation}
V_1(x)=W^2(x)+\frac{dW}{dx}.
\end{equation} 
Note that, since $E=0$ falls in an energy gap, according to the Bloch-Floquet theorem there are two linearly-independent solutions to the equation $H_1 \chi=0$ given by (see, for instance, \cite{Bloch}) $\chi_1(x)=u_1(x) \exp( \mu x)$ and $\chi_2(x)= u_2(x) \exp(-\mu x)$, \revision {where $\mu=\mu_R+i \mu_I$ is the Floquet exponent, $\mu_R={\rm Re}  ( \mu ) >0$, $\mu_I={\rm Im}( \mu)=0, \pi/a$ \cite{noter}}, and $u_{1,2}(x+a)=u_{1,2}(x)$ are periodic functions. In the following analysis we will assume $\chi(x)=\chi_1(x)$, however similar results would be obtained by setting $\chi (x)=\chi_2(x)$. Let us then consider the Hamiltonian $\hat{H}_{2}= \hat{A} \hat{A}^{\dag}=-d^2/dx^2+V_2(x)$, obtained from $\hat{H}_1$ by interchanging the order of the operators $\hat{A}$ and $\hat{A}^{\dag}$. Then it can be readily shown that the potential $V_2(x)$ is given by 
\begin{equation}
V_2(x)=W^2(x)-\frac{dW}{dx}=V_1(x)-2 \frac{dW}{dx}.
\end{equation}
Note that, since $\chi(x)=u_1(x) \exp (\mu x)$,  from Eq.(4) one obtains for the potential $V_2(x)$ the following expression 
\begin{equation}
V_2(x)=V_1(x)-2 \frac{u_1 (d^2u_1/dx^2)-(du_1/dx)^2}{u_1^2}
\end{equation}
which is periodic with the same lattice period $a$ of $V_1(x)$ \cite{note0}. It can then readily proven that the lattices $V_1(x)$ and $V_2(x)$ are isospectral. In fact, let us
indicate by $\psi(x)$ a Bloch eigenfunction of $\hat{H}_1$ with energy $E \neq 0$ belonging to the continuous spectrum of $\hat{H}_1$, i.e. $\hat{A}^{\dag} \hat{A} \psi(x)=E \psi(x)$. It then follows $ \hat{A} \hat{A}^{\dag} \hat{A} \psi(x)=E \hat{A}\psi(x)$, i.e. $\hat{H}_2 \phi(x)=E \phi(x)$ with
\begin{equation}
\phi(x)= \hat{A} \psi(x)=-\frac{d \psi}{dx}+W(x) \psi (x).
\end{equation}
Since $\phi(x)$ - like $\psi(x)$- does not diverge as $ x \rightarrow \pm \infty$, $E$ belongs to the continuous spectrum of $\hat{H}_2$ as well.  Note that Eq.(6) can be inverted, yielding
\begin{equation}
\psi(x)= \frac{1}{E} \hat{A}^{\dag} \phi(x)=\frac{1}{E} \left[ \frac{d \phi}{dx}+W(x) \phi (x) \right].
\end{equation}
In a similar way, if $\phi(x)$ is a Bloch-Floquet eigenfunction of $\hat{H}_2=\hat{A} \hat{A}^{\dag}$ belonging to the continuous spectrum with energy $E \neq 0$, then $\psi(x)$ defined by Eq.(7) is an eigenfunction of $\hat{H}_1$ with the same energy $E$, i.e. $E$ belongs to the continuous spectrum of $\hat{H}_1$ as well. This proves that $\hat{H}_1$ and $\hat{H}_2$ are isospectral, apart from the energy $E=0$, which might belong to the continuous spectrum of $\hat{H}_2$ (but not of $\hat{H}_1$ for construction). To determine whether $E=0$ belongs or not to the continuous spectrum of $\hat{H}_2$, let us notice that $\hat{A}^{\dag} (1/ \chi(x))=-(d \chi /dx)(1/ \chi^2)+W(x) (1 / \chi)=0$, and hence $\hat{H}_2 ( 1 / \chi)=\hat{A} \hat{A}^{\dag} (1/ \chi)=0$. This means that the function
\begin{equation}
\rho(x)=\frac{1}{\chi(x)}= \frac{1}{u_1(x)} \exp(- \mu x)
\end{equation}
is a solution to the equation $H_{2} \phi =0$. The other linearly-independent solution to the same equation is given by $\rho_2(x)=(1/ \chi) \int_0^x dt \chi^2(t)$. Since $\rho(x)$ and $\rho_2(x)$ are unbounded functions as $ x \rightarrow \infty$, it follows that $E=0$ does not belong to the continuous spectrum of $\hat{H}_2$, i.e. the lattices $V_1(x)$ and $V_2(x)$ are isospectral. Finally, note that the superpotential $W(x)$, defined by Eq.(2), can be also written as
\begin{equation}
W(x)=-\frac{d}{dx} {\rm ln} \; [\rho(x)]=-\frac{1}{\rho(x)} \frac{d \rho}{dx}
\end{equation}
where $\rho(x)=1/ \chi(x)$ satisfies the equation $\hat{H}_2 \rho(x)=0$.

\section{ Transparent interface between isospectral crystals} Let us consider two periodic and isospectral lattices with potentials $V_1(x)$ and $V_2(x)$, and let us consider the simplest interface connecting them and described by the potential $V_3(x)=V_1(x)$ for $x<0$ and $V_3(x)=V_2(x)$ for $x>0$. One might naively think that, as the two potentials $V_1(x)$ and $V_2(x)$ are isospectral, the interface should be transparent to any Bloch wave packet propagating across the interface. Unfortunately, this is not the case and a Bloch wave packet rather generally undergoes partial reflection at the interface, in spite of the isospectrality of the two lattices. The reason thereof is that to satisfy the matching conditions at the interface $x=0$, requiring the continuity of the wave function and of its first derivative, the wave at $x<0$ should rather generally include both forward propagating and backward propagating Bloch waves. To realize a transparent interface between two isospectral crystals, we should extend the susy technique for periodic potentials discussed in the previous section. \revision {Following the analysis of Ref.\cite{susy1}}, let us notice that the superpotential is not uniquely defined. Indeed, the Hamiltonian $\hat{H}_2= -d^2/dx^2+V_2(x)$ can be rather generally factorized as  $\hat{H}_2= \hat{B} \hat{B}^{\dag}$, where  $\hat{B}=-d/dx+ Q(x)$ and the superpotential $Q(x)$ is obtained from an arbitrarily chosen solution $f(x)$ to the equation $\hat{H}_2 f=0$ via the relation $Q(x)=-(1/f)(df/dx)$. Note that, if we assume $f(x)=\rho(x)=1/ \chi(x)$, according to Eq.(9) one obtains $Q(x)=W(x)$ and hence $\hat{B}=\hat{A}$. Let us now observe that the most general solution to the equation $\hat{H}_2 f=0$ is given by
\begin{equation}
f(x)=\alpha \rho(x)+ \rho_2(x)=\frac{\alpha+\int_0^x dt \chi^2(t)}{\chi(x)}
\end{equation}  
where $\alpha$ is an arbitrary constant. With the associated superpotential $Q(x)=-(1/f)(df/dx)$, we can construct a third lattice with the Hamiltonian 
\begin{equation}
\hat{H}_3={\hat B}^{\dag } \hat{B}=-\frac{d^2}{dx^2}+V_3(x)
\end{equation}
which is obtained from $\hat{H}_2$ by interchanging the operators $\hat{B}$ and $\hat{B}^{\dag}$.  The potential $V_3(x)$ is obtained from the superpotential $Q(x)$ via the relation
\begin{equation}
V_3(x)=Q^2(x)+\frac{dQ}{dx}.
\end{equation}
Using Eq.(10) with $\chi(x)=u_1(x) \exp(\mu x)$ and the relation $Q(x)=-(1/f)(df/dx)$, after some straightforward calculations one obtains the following expression for the potential $V_3(x)$
\begin{equation}
V_3(x)=V_1(x)+F(x)
\end{equation}
where we have set
\begin{equation}
F(x)=\frac{2\chi^4(x)}{\left[ \alpha+\int_0^x dt \chi^2(t) \right]^2}- \left( \frac{d \chi}{dx} \right)  \frac{4 \chi (x) }{\alpha+\int_0^x dt \chi^2(t) }
\end{equation}
Note that, provided that $\alpha$ is chosen to satisfy the constraint $\alpha > \alpha_0$, with
\begin{equation}
\alpha_0 = \int_{- \infty}^{0} dx \; \chi^2(x) = \int_{- \infty}^{0} dx \;  u_1^2(x) \exp(2 \mu x) < \infty
\end{equation}
$F(x)$ is a bounded and non-singular function, with $F(x) \rightarrow 0 $ as $ x \rightarrow - \infty$ and $F(x) \rightarrow F_0(x)$ with $F_0(x+a)=F_0(x)$ as $ x \rightarrow + \infty$. Therefore, the potential $V_3(x)$, defined by Eqs.(13) and (14), is not periodic and describes an interface separating the two locally-periodic potentials $V_1(x)$ (at $x \rightarrow - \infty$) and $V_1(x)+F_0(x)$ (at $ x \rightarrow \infty$) by a domain wall. Note that, while $F_0(x)$ does not depend on the parameter $\alpha$, the domain wall, i.e. the behavior of $V_3(x)$ near the interface region $x \sim 0$ does depend. Since the parameter $\alpha$ can be chosen arbitrarily, with the solely constraint $ \alpha > \alpha_0$, it follows that there is a one-parameter family of domain walls connecting the two periodic and isospectral lattices $V_1(x)$ and $V_1(x)+F_0(x)$. The main result is that the domain walls synthesized in this way turn out to be fully transparent. In fact, let us indicate by $\psi(x)=u(x) \exp(ikx)$ a forward-propagating Bloch wave of the potential $V_1(x)$ with wave number (quasi-momentum) $k$ and energy $E \neq 0$, belonging to a given allowed band of the lattice, i.e. $\hat{A}^{\dag} \hat{A} \psi(x)=E  \psi(x)$. Taking into account that $\hat{A} \hat{A}^{\dag}=\hat{B} \hat{B}^{\dag}$, it then follows that $\hat{B}  \hat{B}^{\dag}  \hat{A} \psi(x)= E \hat{A} \psi (x)$, and hence  $ \hat{B}^{\dag}  \hat{B}  \hat{B}^{\dag}  \hat{A} \psi(x)= E \hat{B}^{\dag}  \hat{A} \psi (x)$. Since $\hat{H}_3=\hat{B}^{\dag}  \hat{B}$, one obtains that $\hat{H}_3 g(x)=Eg(x)$ with
\begin{equation}
g(x)=\frac{1}{E} \hat{B}^{\dag}  \hat{A} \psi (x)= \frac{1}{E} \left( \frac{d}{dx}+Q(x) \right) \left(- \frac{d}{dx}+W(x)  \right) \psi(x).
\end{equation}

\begin{figure}
\onefigure[width=8cm]{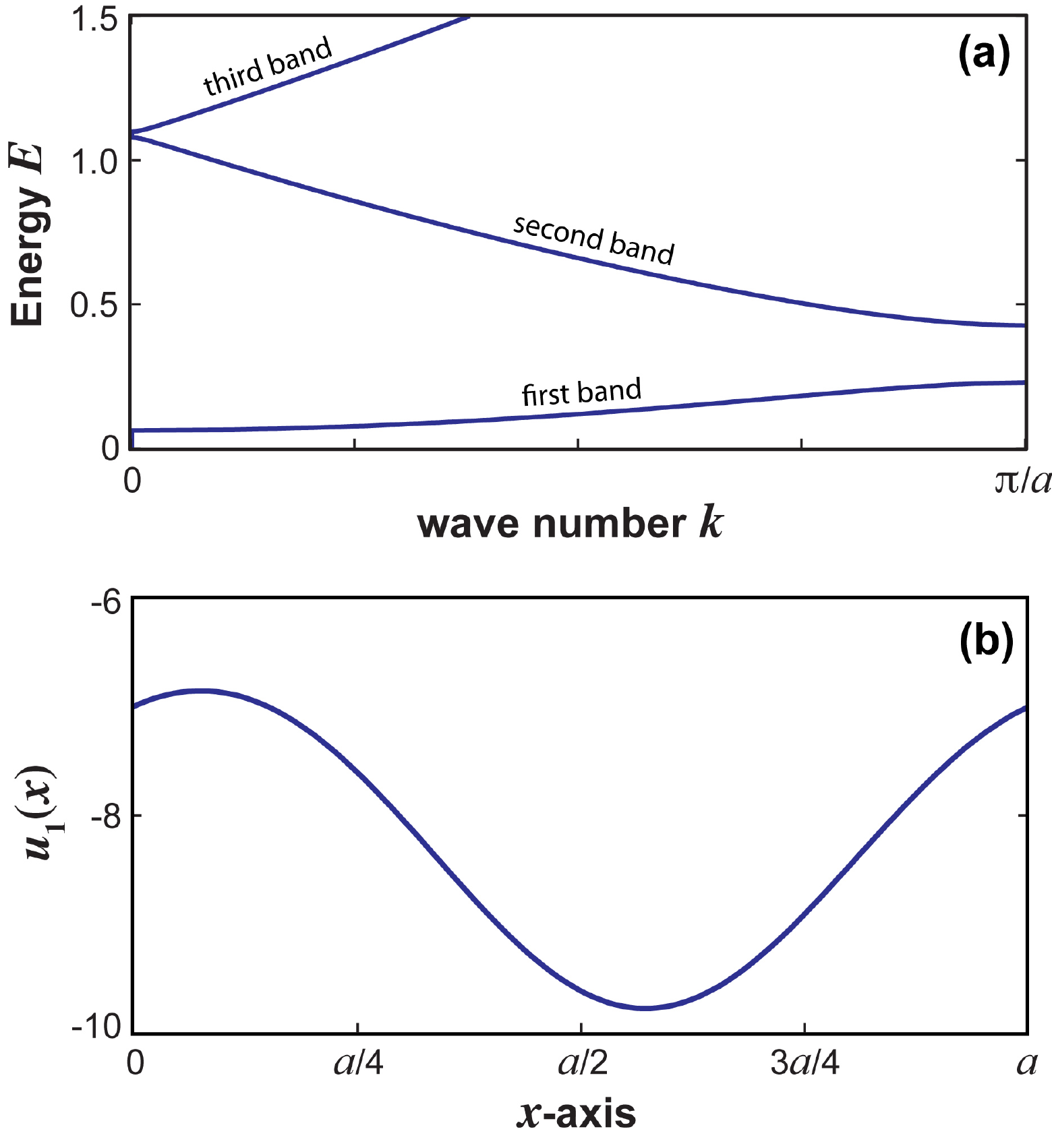}
\caption{(Color online)(a) Band diagram of the Mathieu potential (21) for parameter values $V_0=0.2$, $a=2 \pi$ and $E_0=-0.0818$. (b) Behavior of the periodic part $u_1(x)$ of the evanescent wave $\chi(x)=u_1(x) \exp(\mu x)$ at the energy $E=0$ in a forbidden energy region. The Floquet exponent $\mu$ is $\mu=0.2572$.}
\end{figure}

\begin{figure}
\onefigure[width=8cm]{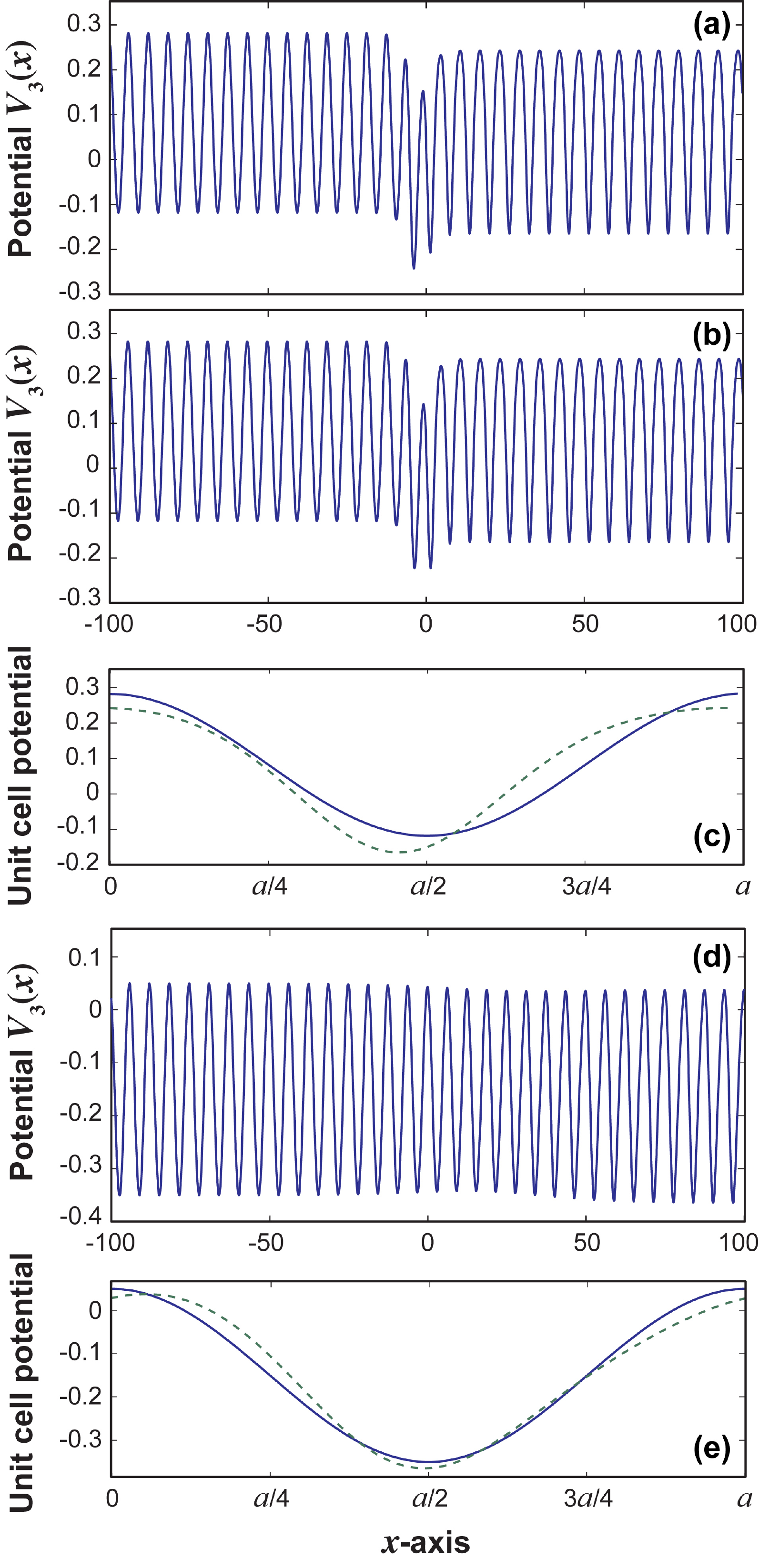}
\caption{(Color online) Examples of transparent domain walls connecting the Mathieu potential $V_1(x)$ with the non-sinusoidal potential $V_1(x)+F_0(x)$, as given by Eqs.(13) and (14), for (a) $\alpha=150$, and (b) $\alpha=170$. The other parameter values are as in Fig.1. In (c) the unit cells of the two periodic lattices $V_1(x)$ (solid curve) and $V_1(x)+F_0(x)$ (dotted curve) are depicted. \revision{ (d) and (e): same as (b) and (c), but for $E_0=0.1503$ and $\alpha=0.1503$}.} 
\end{figure}

\begin{figure}
\onefigure[width=8cm]{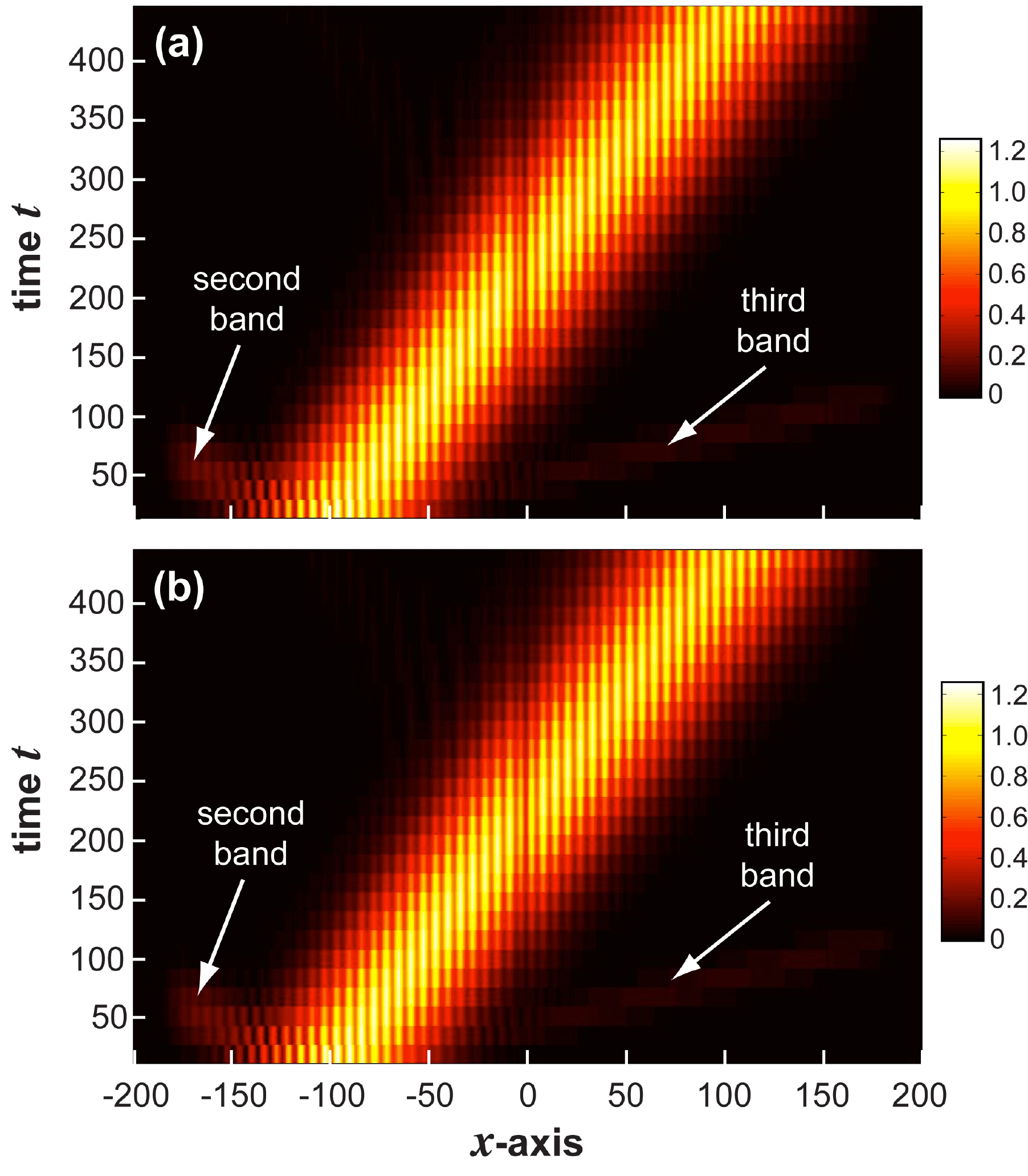}
\caption{(Color online) Numerically-computed evolution of a Gaussian wave packet in the transparent interfaces of Fig.2(a) [upper panel (a)] and 2(b) [lower panel (b)]. The figures show the evolution of $|\psi(x,t)|$ for the initial condition $\psi(x,0)= \exp[-(x-x_0)^2/w^2] \exp(i k_0x)$ for parameter values  $x_0=-100$, $w=40$ and $k_0=1/4$. The initial wave packet mainly excites the lowest band of the lattice,  however small excitation of the two higher bands is visible.} 
\end{figure}

Equation (16) shows that $g(x)$ is an eigenfunction of the lattice $V_3(x)$ with energy $E$. Taking into account that $Q(x)=-(1/f)(df/dx)$ and $W(x)=(1 / \chi) (d \chi /dx)$, from Eqs.(10) and (16) one obtains
\begin{equation}
g(x)= \psi(x)+\frac{1}{E}\frac{\chi(x)}{\alpha+ \int_0^x dt \; \chi^2(t)} \left[ \chi(x) \frac{d \psi}{dx} - \psi(x) \frac{d \chi}{dx} \right]
\end{equation}
with $\chi(x)=u_1(x) \exp( \mu x)$. The transparency of the interface readily follows from the asymptotic behavior of $g(x)$ as $ x \rightarrow \pm \infty$. In fact, as $ x \rightarrow - \infty$ one has $g(x) \simeq \psi(x)=u(x) \exp(i kx)$ whereas $g(x) \simeq h(x) \exp(i kx)$ as  $ x \rightarrow + \infty$, where
\begin{equation}
h(x)=u(x)+ \frac{(ik-\mu)u_1^2 u +u_1^2 (du/dx)-uu_1(du_1/dx)}{E s(x)}
\end{equation}
and 
\begin{equation}
s(x)= \lim_{x \rightarrow + \infty} \exp(-2 \mu x) \int_0^x dt \; u_1^2(t) \exp(2 \mu t).
\end{equation}
Note that $s(x)$, and hence $h(x)$, is a periodic function of $x$ with period $a$.  This means that, as $ x \rightarrow \pm \infty$,  $g(x)$ is of Bloch-Floquet type with the same wave number $k$, indicating the absence of reflected waves.\\ 
Finally, it should be noted that the transparent interface, connecting the two locally-periodic potentials $V_1(x)$ and $V_1(x)+F_0(x)$, sustains a localized (surface) state near the domain wall corresponding to the energy $E=0$ in the gap. The surface state is simply given by
\begin{equation}
g_s(x)=\frac{1}{f(x)}= \frac{\chi(x)}{\alpha+\int_0^x dt \; \chi^2(t)}
\end{equation}
which is a normalizable state, exponentially decaying at $ x \rightarrow \pm \infty$ and satisfying the equation $\hat{H}_3 g_s(x)=0$.

\section{An example of reflectionless interface}
As an example, let us consider the sinusoidal (Mathieu) potential
\begin{equation}
V_1(x)=V_0 \cos( 2 \pi x/a)-E_0
\end{equation}
where $V_0$ is the lattice amplitude and $E_0$ is an energy offset. The Mathieu potential is a well-studied and soluble problem in energy band theory \cite{Sla}. The band structure of the periodic lattice, numerically computed by standard methods, is shown in Fig.1(a) for $a=2 \pi$, $V_0=0.2$ and for an energy offset $E_0=-0.0818$. For such an energy offset, the energy $E=0$ falls in a forbidden region, namely below the lowest lattice band; at this energy the Floquet exponent $\mu$ is real-valued and given by $\mu \simeq 0.2572$. The behavior of the periodic part $u_1(x)$ of the evanescent wave function $\chi(x)=\chi_1(x)=u_1(x) \exp( \mu x)$ is depicted in Fig.1(b). The corresponding critical value $\alpha_0$ of the parameter $\alpha$, defined by Eq.(15), turns out to be $\alpha_0 \simeq 117.45$.  Figures 2(a) and 2(b) show, as an example, the behavior of the potential $V_3(x)$, given by Eqs.(13) and (14), that connects the two periodic lattices $V_1(x)$ and $V_1(x)+F_0(x)$ by a transparent domain wall for $\alpha=150$ [Fig.2(a)] and $\alpha=170$ [Fig.2(b)]. The unit cells of the two periodic lattices  $V_1(x)$ and $V_1(x)+F_0(x)$, that are connected at the interface by the domain wall, are depicted in Fig.2(c). \revision{We note that transparent domain walls could be synthesized by shifting the energy offset $E_0$ in Eq.(21) so that the energy $E=0$ falls inside an energy gap embedded between two transmission bands. As an example, in Fig.2(d) we plot the shape the potential $V_3(x)$ with an embedded transparent domain wall for $E_0=0.1503$ and $\alpha=10$. For such an energy offset, $E=0$ falls in the gap between the first and second lattice bands. The Floquet exponent $\mu$ is given by $\mu=0.0334-0.5i$. The unit-cell behavior of the two isospectral lattices $V_1(x)$ and $V_1(x) + F_0(x)$ that are connected by the domain wall are shown in Fig.2(e). Note that, as compared to the potentials of Figs.2(a-c), in this case the domain wall is smoother and the two isospectral potentials are more similar each other. If the offset $E_0$ in Eq.(21) is further increased to push the energy $E=0$ in a narrower gap between two higher bands, the two isospectral potentials turn out to be almost indistinguishable.}  
\par
We have checked the transparency of the interfaces by direct numerical simulations of the time-dependent Schr\"{o}dinger equation 
\begin{equation}
i \partial_{t} \psi(x,t)=-\frac{\partial^{2} \psi(x,t)}{\partial x^2}+V_{3}(x) \psi(x,t) 
\end{equation}
 using an accurate pseudospectral split-step method.  As an initial condition, we assumed a Gaussian wave packet, of width $w$ and localized at the position $x_0<0$ far from the interface, with a mean momentum $k_0$, i.e. $ \psi(x,0)=\exp[-(x-x_0)^2/w^2] \exp(ik_0x)$.
The input wave packet generally excites Bloch-Floquet states belonging to different bands of the lattice, with a weight that is determined by the so-called Bloch excitation functions  \cite{kepalle}. The group velocities and group velocity dispersion of the wave packet components belonging to the various lattice bands are determined by the derivatives of the band dispersion curves at $k=k_0$. Hence,  the initial wave packet generally breaks into several wave packets that propagate at different speeds and undergo different spreading \cite{kepalle}.  An example of wave packet transmission across the interfaces of Figs.2(a) and (b) is shown in Figs.3(a) and (b), respectively, which depict the evolution of $|\psi(x,t)|$ for parameter values $x_0=-100$, $w=40$ and $k_0=1/4$.  The initial wave packet mainly excites Bloch states belonging to the lowest lattice band, however wave packet components belonging to the second and third lattice bands are visible as well. The figures clearly show that the interfaces connecting the two locally-periodic crystals $V_1(x)$ and $V_1(x)+F_0(x)$ are transparent, i.e. no reflection of the incident wave packet is observed. According to the theoretical analysis, the transparent interface sustains a localized (surface) state, which is given by Eq.(20). In our numerical simulations, this is clearly shown in Figs.4(a) and 4(b), which depict the evolution of $|\psi(x,t)|$ in the lattices of Figs.2(a) and (b) for an initial Gaussian input wave packet $\psi(x,0)$ corresponding to $x_0=0$, $w=10$ and $k_0=0$.\\

\begin{figure}
\onefigure[width=8cm]{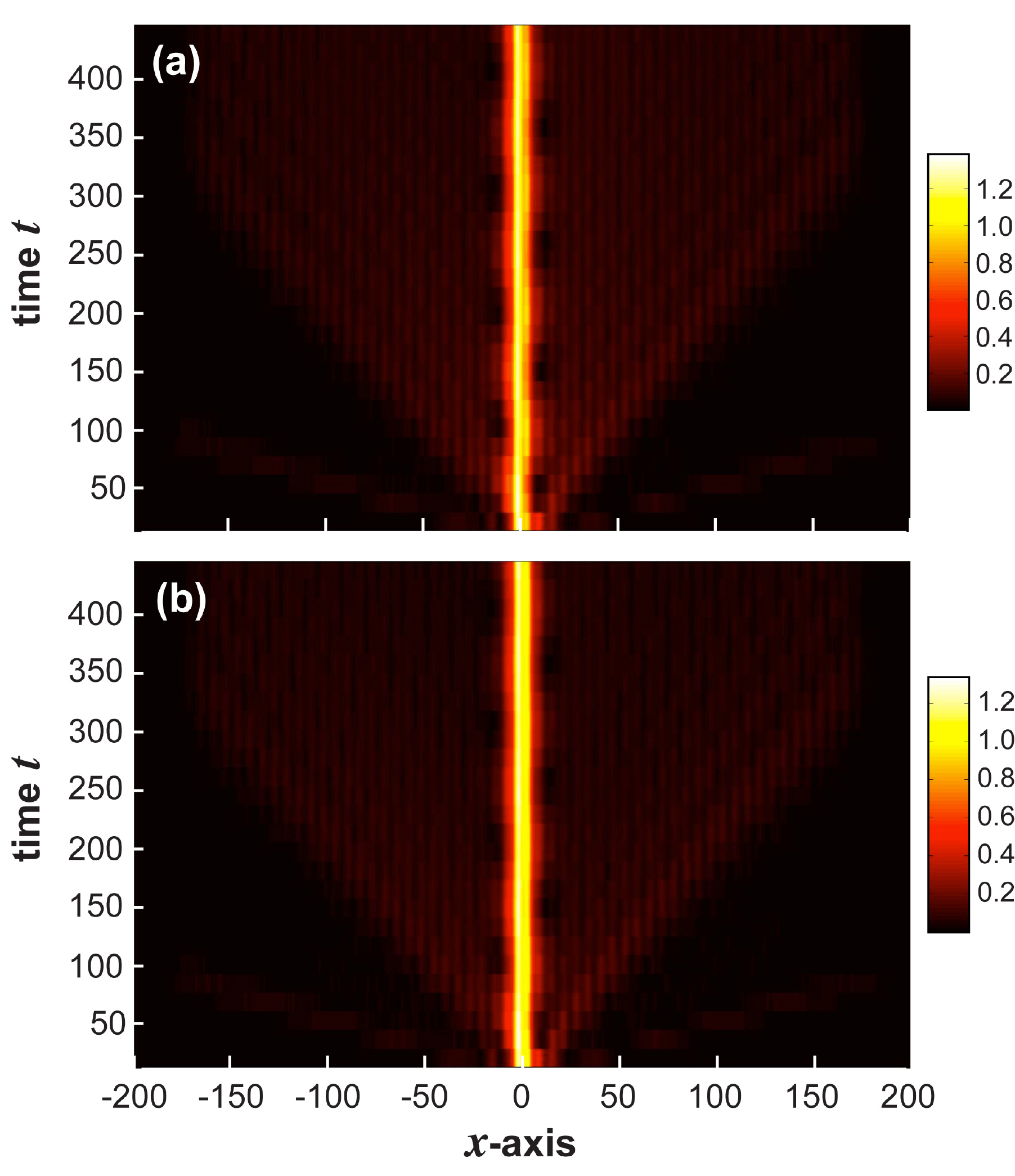}
\caption{(Color online) Same as Fig.3, but for the initial condition $\psi(x,0)= \exp[-(x-x_0)^2/w^2] \exp(i k_0x)$ with  $x_0=0$, $w=10$ and $k_0=0$. In this case excitation of the surface state (20), localized at the interface, is clearly visible.} 
\end{figure}

\section{Conclusions}
Reflection at an interface separating two different media is a rather universal phenomenon which arises because of wave mismatching at the interface. However, in certain cases reflection is suppressed, making the interface transparent. Such an exceptional circumstance occurs, for example, in Klein tunneling of massless Dirac fermions at a potential step or barrier in graphene. Here we have shown that transparent interfaces, connecting two isospectral crystals, can be realized. 
By means of supersymmetric methods of quantum mechanics, we have synthesized a one-parameter family of domain walls, connecting two isospectral crystals, which is fully transparent at any energy of the propagating particle. An example of reflectionless interface has been presented for the sinusoidal (Mathieu) crystal, connected to a non-sinusoidal potential. The transparency of the domain walls has been verified by direct numerical simulations of the time-dependent Schr\"{o}dinger equation. It is envisaged that our results could be of interest to both scattering of matter or optical waves at the interface between periodic optical potentials. The realization of transparent interfaces requires to synthesize special potentials. For cold atoms, arbitrary prescribed
optical potentials can be realized by tailoring the laser intensity profile using
holographic optical beam shapers \cite{holo1,holo2}. For optical waves, artificial media with an arbitrarily-shaped
refractive index profile can be realized by etching subwavelength
structures into a homogeneous high-index dielectric slab, the local effective refractive index being controlled by the duty
cycle of the subwavelength structures \cite{opt1,opt2}.

%\acknowledgments


\begin{thebibliography}{0}
\bibitem{note0}
In fact, if the two crystals have a different band structure, there would exist an energy value $E$ corresponding to propagative waves (i.e. inside a band)  in one crystal and to evanescent waves (i.e. in a gap) in the other crystal. In this case total (Bragg) reflection would be observed at the interface between the two crystals.
\bibitem{note}
We assume that the two potentials $V_1$ and $V_2$ are not related  by a simple translational ($ x \rightarrow x+ \alpha$) or inversion ($ x \rightarrow -x$) transformation. In such cases the two crystals would be trivially isospectral.
\end{thebibliography}
\end{document}